\documentclass{emulateapj}

\newcommand{\mysrc}{CXOU~J010043.1--721134}

\shorttitle{X-Ray Observations of AXP Candidate CXOU~J010043.1--721134}
\shortauthors{McGarry et al.}

\begin{document}
\title{X-Ray Timing, Spectroscopy and Photometry of the Anomalous X-Ray Pulsar
  Candidate CXOU~J010043.1--721134}
\author{M. B. McGarry and B. M. Gaensler\altaffilmark{1}}

\author{S. M. Ransom\altaffilmark{2,3} and V. M. Kaspi\altaffilmark{2}}

\author{S. Veljkovik\altaffilmark{4}}

\altaffiltext{1}{Harvard-Smithsonian Center for Astrophysics, 60 Garden Street, Cambridge, MA 02138}

\altaffiltext{2}{Department of Physics, McGill University, Rutherford Physics Building, 3600 University Street, Montreal, QC H3A 2T8, Canada}

\altaffiltext{3}{Current Address: National Radio Astronomy Observatory,
 520 Edgemont Road, Charlottesville, VA 22903}

\altaffiltext{4}{Harvard College, Cambridge, MA 02138}

\begin{abstract}
We present new X-ray timing and spectral results on the 8.0-second
X-ray pulsar CXOU~J010043.1--721134 from a series of observations
using the {\em Chandra X-ray Observatory}. We find a spin period in
2004~January of $8.020392\pm0.000009$~seconds. Comparison of this to
2001 {\em Chandra}\ observations implies a period derivative $\dot{P} =
(1.88 \pm 0.08) \times 10^{-11}$ s~s$^{-1}$, leading to an inferred dipole surface
magnetic field of $3.9\times10^{14}$~G.  The spectrum
is well fit to an absorbed blackbody of temperature
$kT = 0.38\pm0.02$~keV with a power law tail of photon index $\Gamma =
2.0\pm0.6$. We find that the source has an unabsorbed X-ray flux
(0.5--10~keV) of $4^{+2}_{-1} \times 10^{-13}$~erg~cm$^{-2}$~s$^{-1}$ and a
corresponding X-ray luminosity of $\sim2\times10^{35}$~erg~s$^{-1}$ for
a distance of 60~kpc. These properties support classification of
CXOU~J010043.1--721134 as the seventh confirmed anomalous X-ray pulsar,
the eleventh confirmed magnetar, and the first magnetar to be identified
in the Small Magellanic Cloud.
\end{abstract}

\keywords{pulsars: general --- pulsars: individual(\objectname{\mysrc}) --- stars: neutron}

\section{Introduction}

Before the mid-1990s, it was thought that young neutron stars had short
(P $\ll$ 1s) spin periods, surface magnetic fields B$\sim10^{12}$ G,
and exhibited radio pulsations.  However, recent theoretical and
observational work has revised this picture.  We now
know that a small population of sources known as soft gamma
repeaters (SGRs) are young neutron stars with extreme (B$\sim10^{14-15}$ G) surface magnetic fields, or ``magnetars''.  Another class of objects, anomalous X-ray pulsars (AXPs), are thought to be closely related to SGRs.  Anomalous X-ray pulsars
are so-named due to their high X-ray
  luminosities and unusually fast spindown rates, with no evidence of
  variation due to binary motion, which distinguish them from both
  isolated radio pulsars and accreting X-ray binaries.  Most significantly, AXPs exhibit a
  small range of spin periods (5-12
  seconds) with steady spindown that, if powered by magnetic dipole radiation, imply magnetic fields 2-3
  orders of magnitude larger than those of typical radio pulsars
  \citep{tho96}.  The magnetar model for AXPs is supported by the
  SGR-like bursting activity that has been observed from some of these sources \citep{gav02,kas03}.
  Other characteristics include spectra that are
  well fit by a blackbody model with a temperature $kT\sim0.5$ keV and a power law tail with photon index $\Gamma$=2-4, and lack of a measurable
  binary companion that could otherwise produce similar X-ray luminosities
  through accretion \citep[see][for a summary of the properties of
    known magnetars]{woo05}.  With unabsorbed luminosities in the range of $10^{34}$ to $10^{36}$
  erg s$^{-1}$, inferred magnetic fields of $10^{14}$ - $10^{15}$
  Gauss, and characteristic ages of $\sim8-200$ kyr, it
  has been firmly established that AXPs belong to the magnetar family \citep{gav04}.

 The X-ray source \mysrc\ in the Small Magellanic Cloud (SMC) was first detected by \emph{Einstein}
  in 1979, and was later observed several times by \emph{ROSAT} and \emph{ASCA}, but its
  periodicity was not noticed until an archival search for X-ray
  pulsations was carried out on a 2001 \emph{Chandra}
  observation of a nearby field \citep{lam02}.  The period of this source was determined to be 8.0
  seconds and its  X-ray luminosity was measured to be on the order of $10^{35}$ erg s$^{-1}$, assuming a distance to the
 SMC of 60 kpc \citep{lam02,lam03,maj04}.  Such a high luminosity and
  long spin period suggested that this object may belong to the
  growing class of AXPs \citep{lam02,maj04}. 

In this Letter we present new \emph{Chandra} results supporting \mysrc\
as the newest member of the AXP class.  We present timing analysis
from which we measure a period derivative, spectral results, and an
analysis of archival \emph{Chandra} and \emph{XMM} data that assesses
the flux characteristics of the source over time.

\section{Observational Setup} \label{setup}
   Five new \emph{Chandra} observations of \mysrc\ were completed between 2004 January 27 and
    2004 March 25.  The source
    was clearly detected at each epoch, located at (J2000) right ascension $01^{h}00^{m}43\fs03$, declination
$-72^{\circ}11^{\prime}33\farcs6$, with a radial
uncertainty of $0\farcs5$.   Coordinates from the \emph{Chandra} pointings have not yet been
refined using an optical reference.  Each observation had
    approximately 16 ks of on-source time, and the
    observations were spaced in a geometric series designed for a
    phase-coherent timing analysis.  All five observations
    were made with the ACIS-S3 CCD in Very Faint mode.
    A 1/8 subarray was used to reduce the time resolution to 0.4 s.

    Additionally, we analyzed serendipitous archival observations from both
    \emph{Chandra} and \emph{XMM}.
    The archival 2001 May 15 \emph{Chandra}
    observation was made in faint mode, and the source fell on the ACIS-I1 CCD
    approximately 10$^{\prime}$ from the aimpoint.  We reprocessed the
    level 1 events file and applied the CTI, standard gain, and time-dependent gain
    corrections.  \emph{XMM} archival observations were taken on 2000
    October 10 and on 2001 November 20 using the EPIC MOS detectors, and the 2001
    observation also used
    the EPIC PN detector.  Both datasets lost some observation time due to
    high background.   The
    source location in each \emph{XMM} observations is also off-axis,
    and the source lies on a chip
    gap on the MOS2 instrument.  Therefore, we discarded all MOS2 data in this analysis.  Table \ref{alldatafits} contains
    detailed information on each observation.  We processed \emph{Chandra}
    observations using CIAO 3.1 and CALDB 2.28, and
    \emph{XMM} observations using SAS 6.0.0.

\section{Timing Analysis} \label{timing}

The five 2004 \emph{Chandra} observations of \mysrc\ were separated by
successive intervals of 0.40, 1.72, 8.9, and 46.4 days.  For each
epoch, we extracted the events from a $2^{\prime\prime}$ radius region
centered on the coordinates of the source and corrected their
arrival times to the solar system barycenter.  We measured
pulse profiles for each observation by folding the X-ray events at a
range of periods around the $P$$\sim$8.02\,s spin-period of the pulsar
as reported by \citet{lam02,lam03}.  The profiles with the highest significance (as determined by
maximizing $\chi^2$ with respect to a model with no
pulsations) were cross-correlated and co-added in phase to create an
integrated template pulse profile from all the observations.  We then
determined pulse arrival times by cross-correlating the profiles from
each observation with the high signal-to-noise template profile.
From the three \emph{Chandra} observations in 2004 January, we unambiguously measure a spin
period $P = 8.020392(9)$~s, where the number in parentheses indicates
the uncertainty in the last digit.  Unfortunately, the fractional phase errors from these measurements
were $\sim$0.06, significantly larger than the 0.02$-$0.03 fractional
phase errors that we were expecting.  These larger arrival time
uncertainties mean that the spin period determined above
from the three 2004 January observations was not of sufficient accuracy to
unambiguously account for each rotation of the pulsar between the 2004
January and February observations, let alone during the much longer
gap between the 2004 February and March observations.

From the archival {\em Chandra}\
observation taken on 2001 May 15, we measure $P=8.0188(1)$~s, which
is significantly shorter than $P = 8.020392(9)$~s in 2004~January
as determined above.  If this period difference is due to a relatively constant
spin-down, it implies $\dot P$=1.9(1)$\times$10$^{-11}$ s~s$^{-1}$.  If we then force a phasing of the five 2004 observations to most
closely match the average spin-down between 2001 May and 2004 January, we
determine $\dot P$=1.88(8)$\times$10$^{-11}$ s~s$^{-1}$.  A single phase wrap
between the 2004 February and 2004 March observations implies $\dot
P$=1.26(8)$\times$10$^{-11}$ s~s$^{-1}$ or $\dot
P$=2.50(8)$\times$10$^{-11}$ s~s$^{-1}$, both of which are clearly inconsistent
with the average spin-down.  It is therefore highly
likely that the true spin-down value of this source is $\dot
P$=1.88(8)$\times$10$^{-11}$ s~s$^{-1}$.

\section{Spectroscopy} \label{all_obs}

For the spectral analysis on the 2004 \emph{Chandra} observations, we used an extraction radius of
$2^{\prime\prime}$ for the source region and $29\farcs03$ for the
background region.  For
the 2001 off-axis \emph{Chandra} observation, the source region was an ellipse
with major and minor axes $10\farcs82$ and $9\farcs84$
respectively, and background extraction circle with radius
$92\farcs50$.  In all \emph{Chandra} analyses, the background regions were not centered on
the source but did cover the source and so the source region was excluded.  For the \emph{XMM} observations, the source
extraction circles for the MOS and PN detectors had radii of $12\farcs40$ and
$19\farcs20$, respectively, while the background regions for
both were $65\farcs60$ and did not overlap the source region.

 We successfully fit each spectrum in Sherpa\footnote{http://cxc.harvard.edu/sherpa/} to a model containing an
    absorbed blackbody plus a power law, accounting for absorption
    from the Milky Way and the SMC.   We assumed a Galactic absorbing
    column $N_{H}$ =$ 4.3\times10^{20}$ cm$^{-2}$ \citep{naz03}.  Standard SMC
    elemental abundances were used the hydrogen column due to
    the SMC contribution was allowed to vary\footnote{Abundances for elements Co and Ca were
    not included in the model of \citet{rus92}, and so both were assigned a value of 0.3171 with
    respect to solar in the spectral model based on the average
    abundance of the other elements.  Variations in the Co and Ca
    abundances do not result in significant changes to the fits.} \citep{rus92}.  The temperature and photon index,
    and their normalizations, were allowed to vary in their respective models.

Table~\ref{ch04fits}
    compares various models that we fit to the data.  A blackbody plus
power law model provides a much better fit than the single model fits
that were previously attempted \citep{lam02,naz03,maj04}.  Fitting to a simple absorbed
    power law model yields a poor result, as shown by the
    reduced $\chi^{2}$ = 2.12.  Fitting to a simple absorbed
    blackbody model yields a reduced $\chi^{2}$ = 1.10.  However, this
    fit is infeasible because it requires that there be no absorption column
    contribution from the SMC (${N_H}_{SMC}$).  In fact, if ${N_H}_{SMC}$ is frozen at
    a reasonable value we are still not able to fit a sensible model to
    the data.

  Statistics were
    calculated to 90\% confidence ($\sigma$ = 1.6).  In each dataset,
    we excluded data below 0.5 keV to reflect telescope sensitivity
    and above 4.0 keV because the source spectrum is not detected beyond this.  We subtracted the background, although it was minimal in the
\emph{Chandra} data, and grouped the spectra into bins of at least 25
counts for analysis.

\subsection{\emph{Chandra} 2004 Observations}\label{chandradata}

    We fit the five 2004 \emph{Chandra} datasets simultaneously and
    find that ${N_H}_{SMC}$ = $(3\pm4)\times10^{21}$ cm$^{-2}$.  The large uncertainty in this
    absorbing column is due to the coupling between the absorbing
    column and the power law index $\Gamma$.  We find a blackbody
    temperature \emph{kT} = 0.38$\pm$0.02 keV, power law photon index $\Gamma$ = 2.0$\pm$0.6, and
    reduced $\chi^{2}$ = 0.97.  The
    unabsorbed 0.5-10 keV luminosity of the source in these \emph{Chandra} observations is
    $\sim2\times10^{35}$ erg s$^{-1}$ at a distance of 60 kpc, and the blackbody component
    contributes $\sim$50\% of the luminosity. 

\subsection{2000 and 2001 Archival Observation}\label{archivaldata}
    We also analyzed the 2001 archival \emph{Chandra}
    observation.  The model applied to the previously described \emph{Chandra} data also yields an excellent fit to this observation,
    with ${N_H}_{SMC} = (8\pm6)\times10^{21}$ cm$^{-2}$,
    \emph{kT} = 0.36$\pm$0.03 keV, $\Gamma$ = 1.9$\pm$0.4, and reduced $\chi^{2}$ = 1.15.   The
    unabsorbed 0.5-10 keV
    luminosity of the source is
    $\sim2\times10^{35}$ erg s$^{-1}$, and the blackbody component
    contributes $\sim$60\% of the luminosity.
 It
    should be noted that these data were fit after applying a user-contributed
    time-dependent gain correction.  This software accounts for an instrumental
    effect that has not been fully addressed by the default Sherpa
    package for older observations.

   The best fitting model to the 2000 and 2001 \emph{XMM} data is
    again the absorbed
    blackbody with a power law component.  Though we also fit the two
    \emph{XMM}
    datasets individually, we found that they were well modeled when
    fit jointly.  This model implies ${N_H}_{SMC}$ =
    $(5\pm7)\times10^{21}$ cm$^{-2}$, \emph{kT} = 0.31$\pm$0.04 keV,
    $\Gamma$ = 2.0$\pm$0.5, and reduced $\chi^{2}$ = 1.08.  The
    unabsorbed 0.5-10 keV
    luminosity of the source as seen by \emph{XMM} is
    $\sim2\times10^{35}$ erg s$^{-1}$, and the blackbody component
    contributes $\sim$40\% of the luminosity.

\section{Long Term Flux Behavior} \label{variability}
  We have also examined the flux variability of \mysrc\ over the last
  three decades using the data in \S \ref{all_obs}, plus count rates from archival data reported by
  \citet{lam02}.  We inferred the fluxes and their errors for the older
  observations using the following method.  For both the \emph{Chandra} and
  \emph{XMM} observations presented in this paper, we used Sherpa's eflux command to calculate the flux
  for the best-fit parameters of ${N_H}_{SMC}$,
  $\Gamma$, and \emph{kT}.  We also determined minimum and
  maximum possible fluxes, by applying the errors in the spectral fits, which
  were calculated at the 90\% confidence level, in order to achieve
  the largest possible flux range.  Specifically, the minimum flux
  occurs when ${N_H}_{SMC}$ and $\Gamma$ are at a maximum and \emph{kT} is at a
  minimum, and the maximum flux occurs in the reverse scenario.  We
  applied a correction to account for the correlations in the
  uncertainties of the three parameters and defined the flux error to be the difference between
  the flux defined by the best-fit parameters and these extreme flux
  values.  

\begin{figure}
  \epsscale{.80}
  \plotone{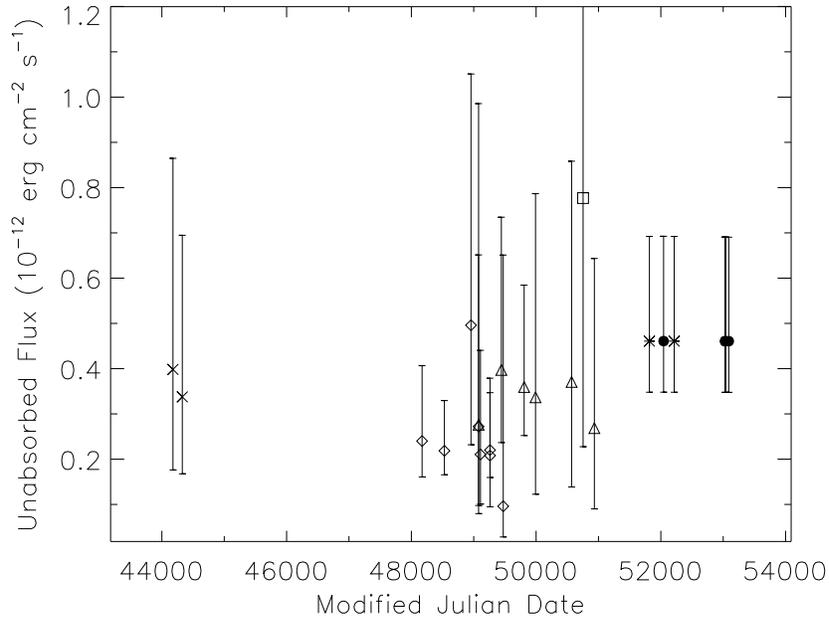}
  \caption{Estimated unabsorbed fluxes for \mysrc\ over the last three
    decades of observations, using the BB+PL model from
    Table~\ref{ch04fits} in the energy range 0.2-10 keV.  Each instrument is denoted with a separate plot symbol as
    follows: X - \emph{Einstein} IPC, $\diamond$ - \emph{ROSAT} PSPC, $\triangle$ - \emph{ROSAT}
    HRI, $\Box$ - \emph{ASCA} GIS, $\ast$ - \emph{XMM} MOS/PN, $\bullet$ - \emph{Chandra} ACIS.  Flux estimates
    from observations presented in this \emph{Letter} were determined by using
    Sherpa's eflux command.  Those not presented in this \emph{Letter} are derived from the
    observed count rates listed in \citet{lam02} by using HEASARC
    WebPIMMS.
    \label{fluxplot}
  }
\end{figure}

Fluxes for the observations from previous X-ray satellites listed in \citet{lam02} could not be calculated in
this manner without reprocessing each dataset, so we developed a
method by which to extrapolate their fluxes and errors.
  We converted the errors from our \emph{Chandra} observations into fractional uncertainties and
  then compared them to the fractional uncertainty in their measured
  count rate to obtain a count rate-flux uncertainty relation.
  Using the best-fit model for the 2004 \emph{Chandra}
  observations (see Table \ref{alldatafits}), we converted from count
  rate to flux using HEASARC's
  WebPIMMS\footnote{http://heasarc.gsfc.nasa.gov/Tools/w3pimms.html}.  We then used our
  count rate-flux relation to extrapolate the
  uncertainties in these older observations.  Results are plotted in Figure \ref{fluxplot}.  Because small changes in the fit parameters lead to large
  variation in flux for a given count rate, it is difficult to compare
  fluxes across multiple epochs.  However, at face value, the flux
  behavior does not seem to demonstrate variation spanning more than one order of
  magnitude over a timescale of $\sim$25 years.

\section{Discussion}

The properties of \mysrc\ are completely consistent with those
expected of an AXP.   In 2004 January, the source had a period of
8.020392(9) seconds and a $\dot{P}=1.88(8)\times10^{-11}$ s~s$^{-1}$.   If this spin-down is due to ``standard'' dipole radiation, the implied
surface magnetic field strength is 3.2$\times$10$^{19}$($P\dot
P$)$^{1/2}$\,G = 3.9$\times$10$^{14}$\,G, consistent with magnetic fields of other AXPs.  Using the estimated period derivative, it is possible to calculate the
characteristic age and spindown
luminosity of the source.  A period derivative
$\dot{P}=1.88(8)\times10^{-11}$ s~s$^{-1}$
implies a characteristic age of $P/2\dot P$ = 6800 years (with a 4\% uncertainty), consistent
with characteristic ages of magnetars \citep{gae01}.  \mysrc\ has a
spin-down luminosity of $4\pi^{2}I\dot P/P^{3} \equiv 1\times10^{33}$
erg s$^{-1}$, where I $\equiv$ 10$^{45}$ g cm$^{2}$ is the assumed moment of
inertia of the neutron star.

The source spectrum is well
described by a photon index of $\Gamma$ = 2.0$\pm$0.6 and a
blackbody temperature of \emph{kT} = 0.38$\pm$0.02 keV.  This implies a
0.5-10 keV unabsorbed luminosity of $\sim$2$\times10^{35}$ 
erg s$^{-1}$, where the blackbody contributes $\sim$50\% of the
luminosity (at a distance of 60 kpc).  This fit is consistent with those of other AXPs,
    which tend to have \emph{kT} $\sim$ 0.5 keV and $\Gamma\sim$2-4.
    Although the blackbody contribution to the luminosity of an AXP
    is highly energy dependent, the value of $\sim$50\% in the case of
    \mysrc\ is consistent with other
    AXPs in the 1-4 keV range \citep{oze01}.

 It has been suggested that 
the source could be a Be X-ray binary with a $\sim$25 day period because there is an
optical Be star within $2^{\prime\prime}$ of the source \citep{naz03}.  X-ray binaries
are also described by a blackbody plus power law model, but generally their
blackbody temperature is higher (i.e. 1-2 keV) \citep{hab04}.
However, our timing analysis rules out a 10-30 day binary due to
the absence of acceleration by a massive companion in the first three \emph{Chandra}
obsevations since the phase connects unambiguously
over $\sim$3 days.  The phase wrap after the third
observation does leave open the possibility of longer period binary
motion, however, the long term period change seems to match the best phase
connected solution, which likely rules out any binary companion and
argues strongly for a relatively steady spin-down.  We further note that \citet{dur05} have identified a likely
optical counterpart to \mysrc\, distinct from the adjacent B star.  This object provides a convincing alternative to the Be
X-ray binary interpretation.

\mysrc\ does not show strong evidence of variability over the timescale of
  years.  Although not all AXPs are variable, some AXPs have been found to vary across one to
  two orders of magnitude \citep{gav04,ibr04}.  X-ray
  binaries also exhibit variations,
  by factors of up to 20 over a similar timescale, and can
  change by a factor
  of more than 100 during giant outbursts \citep{hab04}.  Even
  considering the large uncertainties, we do not see variability on
  this scale in
\mysrc\ over the past 25 years.  While it does not conclusively
  characterize the object, apparent lack of variability across orders
  of magnitude does not challenge the categorization of this source as an AXP.

In summary, the spectral, timing, and photometric properties of \mysrc\ are consistent
with the properties of other AXPs \citep{mer02}, expanding
this class of magnetars to seven confirmed members.  We note that the
source is the only known magnetar in the
SMC.  If the magnetar
birth rate is $\sim$10\% of that of radio pulsars \citep{kou94,gae05},
then for an SMC supernova rate of $(5\pm3)\times10^{-4}$ per year
\citep{cra01} and a magnetar lifetime on the order of $10^{4}$
yr \citep{col00,gae01}, we can expect 0.5 $\pm$ 0.3 magnetars to exist
in the SMC at a given time.  This would be consistent with the fact
that only this magnetar is known in the SMC
thus far, despite several archival searches for X-ray sources
\citep{maj04,hab04,lay05}.  Interestingly, \citep{ibr04} report the discovery of a transient
magnetar, detected only because of an outburst that resulted in a
substantial flux enhancement that faded on a time scale of months.
Similarly, the highly variable 7-s pulsar AX J1845-0258 may be another such
source (Vasisht et al. 2000).  These sources suggest that the
magnetar birthrate could be substantially higher than has been estimated,
possibly even comparable to the radio pulsar birthrate.  Continued monitoring of
the SMC will provide a useful test of these possibilities.  In any case, \mysrc, with a well-known distance and low extinction, will be a
valuable probe into this exotic class of objects.

We thank Fernando Camilo, Cara Rakowski, Michael Garcia, and Terry Gaetz for useful discussions, and the
\emph{Chandra} Helpdesk for software support.  This work was supported
by NASA through contract NAS 8-39073 and SAO grant GO4-5065X.

\begin{deluxetable}{ccccccccc}
\tabletypesize{\tiny}
\tablecaption{Spectral Fits to \emph{Chandra} and \emph{XMM} Datasets\label{alldatafits}}
\tablewidth{0pt}
\tablehead{
\colhead{Dataset \& Obsid} & 
\colhead{Date} &
\colhead{On-Source} &
\colhead{Count Rate} &
\colhead{${N_H}_{SMC}$} & 
\colhead{$\Gamma$} &
\colhead{$\emph{kT}_{BB}$} &
\colhead{$\chi^{2}/DOF$} &\\
\colhead{} & 
\colhead{} &
\colhead{Time (ks)} &
\colhead{($10^{-2}$ ct s$^{-1}$)} &
\colhead{$(10^{21}$ cm$^{-2})$} & 
\colhead{} &
\colhead{(keV)} & 
\colhead{} 
}

\startdata
XMM 0110000201 & 2000 Oct 10 & 14.5 & 5.4(2)  & 1(5) & 1.8(1) &  0.35(6) & 21.2/21  \\
CXO 1881       & 2001 May 15 & 98.7 & 6.03(8) & 8(6) & 1.9(4) &  0.36(3) & 130/113  \\
XMM 0018540101\tablenotemark{a}
               & 2001 Nov 20 & 24.7/20.2 & 4.0(1)/15.1(3)
                                              & 6(8) & 2.0(3) &  0.30(5) & 131/126  \\
CXO 4616       & 2004 Jan 27 & 16.2 & 8.8(5)  & 3(4) & 2.0(6) &  0.38(2) & 45.9/38  \\ 
CXO 4617       & 2004 Jan 28 & 16.1 & 8.8(5)  & 3(4) & 2.0(6) &  0.38(2) & 40.8/39  \\ 
CXO 4618       & 2004 Jan 29 & 16.5 & 8.5(5)  & 3(4) & 2.0(6) &  0.38(2) & 34.3/39  \\ 
CXO 4619       & 2004 Feb 07 & 16.7 & 8.6(5)  & 3(4) & 2.0(6) &  0.38(2) & 35.8/39  \\ 
CXO 4620       & 2004 Mar 25 & 15.6 & 9.2(5)  & 3(4) & 2.0(6) &  0.38(2) & 49.0/37  \\ 
\enddata
\tablecomments{Parentheses
  following each value represent the error in the last digit; ``On-Source
  Time'' refers to the amount of usable time recovered from the
  original observation, excluding deadtime caused by CCD readout,
  high background periods, etc.; Count Rate is in the default energy
  range for each instrument (\emph{Chandra}: 0.2-10keV, \emph{XMM}:
  0.4-10keV); ${N_H}_{SMC}$ = the
  absorbing column due to the Small Magellanic Cloud only, assuming a
  Galactic absorbing column of $4.3\times10^{20}$ cm$^{-2}$; $\Gamma$
  = the power law photon index; $\emph{kT}_{BB}$ = blackbody characteristic
temperature; $\chi^{2}$ value is for the fit of the absorbed black
body plus power law model; The unabsorbed
0.5-10keV flux for the five 2004 \emph{Chandra} observations is
$4^{+2}_{-1} \times 10^{-13}$~erg~cm$^{-2}$~s$^{-1}$ , and in all other
 observations is $5^{+2}_{-1} \times 10^{-13}$ ~erg~cm$^{-2}$~s$^{-1}$.}

\tablenotetext{a}{Values for MOS1 and PN respectively.}
\end{deluxetable}

\begin{deluxetable}{ccccc}
\tabletypesize{\scriptsize}
\tablecaption{Joint Spectral Fits to 2004 \emph{Chandra} Data\label{ch04fits}}
\tablewidth{0pt}
\tablehead{
\colhead{Model} & 
\colhead{${N_H}_{SMC}$(cm$^{-2}$)} &
\colhead{$\Gamma$/$\emph{kT}_{BB}$(keV)} &
\colhead F($10^{-13}$\,erg\,cm$^{2}$\,s$^{-1}$) & 
\colhead{$\chi^{2}/DOF$} 
}

\startdata
BB    & \nodata\tablenotemark{a} & \nodata/0.40(8)  & $\sim$4 & 235/214 = 1.1 \\
PL    & 8(1)$\times10^{21}$      & 2.20(5)/\nodata  & $\sim$6 & 455/214 = 2.1 \\
BB+PL & 3(4)$\times10^{21}$      & 2.0(6)/0.38(2)   & $\sim$4 & 205/212 = 0.97 \\ 
\enddata
\tablecomments{Parentheses
  following each value represent the error in the last digit; $N_H$ = the
  absorbing column due to the Small Magellanic Cloud only, assuming a
  Galactic absorbing column of $4.3\times10^{20}$ cm$^{-2}$;  $\Gamma$
  = the power law photon index; $\emph{kT}_{BB}$ = blackbody characteristic
temperature; F = Unabsorbed flux for the model in the 0.5-10
keV range.}
\tablenotetext{a}{The data could not be fit within Sherpa's
  limits for ${N_H}_{SMC}$.}

\end{deluxetable}

\end{document}